\documentclass[prd,nofootinbib,preprint,superscriptaddress]{revtex4}

\usepackage{amsmath, amssymb, amsthm, graphicx, epsfig, fancyhdr,epsfig, slashed, mathrsfs}
\usepackage{ulem}
\usepackage{tikzsymbols}
\usepackage{natbib}
\usepackage{float}
\usepackage{xcolor}

\usepackage[utf8]{inputenc}

\DeclareUnicodeCharacter{2212}{-}

\usepackage[bookmarks=true,bookmarksopen=true,
                    bookmarksnumbered=true,
                    colorlinks=true,linkcolor=red,
                    citecolor=red]{hyperref}
                    
\definecolor{lime}{HTML}{A6CE39}
\DeclareRobustCommand{\orcidicon}{
	\begin{tikzpicture}
		\draw[lime, fill=lime] (0,0) 
		circle [radius=0.2] 
		node[white] {{\fontfamily{qag}\selectfont \tiny ID}};
		\draw[white, fill=white] (-0.0625,0.095) 
		circle [radius=0.007];
	\end{tikzpicture}
	\hspace{-2mm}
}

\foreach \x in {A, ..., Z}{\expandafter\xdef\csname orcid\x\endcsname{\noexpand\href{https://orcid.org/\csname orcidauthor\x\endcsname}
		{\noexpand\orcidicon}}
}

\newcommand{\be}{\begin{equation}}
	\newcommand{\ee}{\end{equation}}
\newcommand{\bea}{\begin{eqnarray}}
	\newcommand{\eea}{\end{eqnarray}}

\newcommand{\beq}{\begin{equation}}
	\newcommand{\eeq}{\end{equation}}

\def\nn{\nonumber}

\def\ra{\rightarrow}
\def\mpl{M_{\rm Pl}}

\begin{document}
\hspace*{\fill} HRI-RECAPP-2022-012

	\title{Origin of neutrino masses, dark matter, leptogenesis, and inflation in a seesaw model with triplets}

	\author{Pritam Das\orcidA{}}
	\email{prtmdas9@iitg.ac.in}
	\affiliation{Indian Institute of Technology Guwahati, Assam, India-781039}

	\author{Najimuddin Khan\orcidC{}}
	\email{najimuddinkhan@hri.res.in}
	\affiliation{ Harish-Chandra Research Institute, A CI of Homi Bhabha National Institute, Chhatnag Road, Jhunsi, Prayagraj 211019, India}

	\begin{abstract}
		We consider a new physics model, where the Standard Model (SM) is extended by hyperchargeless $Y=0$ triplet fermions and Higgs triplet with hypercharge $Y=2$. The first two generation fermion triplets are even under the $Z_2$ transformation. In contrast, the third fermion triplet and scalar triplet are odd under the same $Z_2$ transformation. It is a unifying framework for the simultaneous explanation of neutrino mass and mixing, dark matter, baryogenesis, inflation, and reheating temperature of the Universe. The two $Z_2$ even neutral fermions explain the neutrino low energy variables, whereas the third one can serve as a viable dark matter candidate, explaining the exact relic density. The scalar triplet is coupled nonminimally to gravity and forms the inflaton. We calculate the inflationary parameters and find them consistent with the new Planck-2018 constraints. We also do the reheating analysis for the inflaton decays/annihilations to relativistic SM particles. The triplet fermions associated with $Z_2$ even sector can provide the observed baryon asymmetry of the Universe at the TeV scale.

	\end{abstract}
	
	\maketitle
	
	\section{Introduction}
	\label{sec:intro}
The confirmation of a Higgs boson~\cite{ATLAS:2012yve, CMS:2012qbp, Giardino:2013bma} at mass $\sim$125.5 GeV has solidified the mechanism of Electroweak Symmetry Breaking (EWSB). Ten years have passed since the discovery of the Higgs boson at the Large Hadron Collider (LHC) by the ATLAS-CMS collaborations. The precision Higgs measurements
which followed all agree with the Standard Model (SM) predictions.
In the meantime, the two ATLAS and CMS collaborations have searched for many hypothetical new physics particles. This tremendous effort was so far failed, and
confirming, once again, the SM, which put the theoretical, as well as experimental physics community in puzzle state.	
Different theoretical incompleteness, such as the hierarchy problem related to the Higgs mass, mass hierarchy and mixing patterns in the quark and leptonic sectors, suggest the existence of new physics beyond the SM. 
Again various earth-satellite-based experimental observations, such as the non-zero neutrino mass, the mysterious nature of dark matter (DM) and dark energy, the baryon-antibaryon asymmetry, and inflation in the early Universe indicate the existence of new physics.

The experimental results of the neutrino oscillation phenomenon are strong evidence to have new physics in addition to SM. The oscillation experiments~\cite{Abe:2016nxk, An:2012eh, Abe:2011fz} of atmospheric, solar, reactor and accelerator neutrinos predicted that the three (at most four) flavours of neutrinos mix and have a tiny mass.
These oscillation experiments can predict the mixing angles ( $\theta_{ij}$) and mass square differences ($\Delta m_{ij}^2=m_i^2-m_j^2$) only. The absolute mass of the individual neutrino mass eigenstates are still not known.
From other observations, we have a stringent constraint on the sum of all neutrino mass eigenvalues ( $\sum_{m_i}<0.117$ eV~\cite{Choudhury:2018byy}, with $i=1,2,3$).	
The measurements of cosmic microwave background (CMB) anisotropies, cosmology-based experiments such as  WMAP~\cite{Bennett:2012zja} and PLANCK~\cite{Aghanim:2018eyx} have suggested the existence of an unknown, non-baryonic and non-luminous component of matter. It is known as dark matter (DM).	
Astrophysical evidence such as galaxy cluster observations by F. Zwicky~\cite{Zwicky:1937zza}, galaxy rotation curves~\cite{Freese:2008cz}, bullet cluster~\cite{Clowe:2006eq} also agrees with the presence of DM. Now, one can explain the dark matter relic density as $\Omega h^2=0.1198\pm 0.0012$~\cite{Aghanim:2018eyx} using various theories~\cite{Kolb:1990vq, Hall:2009bx}.	
The super-horizon anisotropies in the CMB data~\cite{Bennett:2012zja,Aghanim:2018eyx} have also suggested that the early Universe underwent a period of rapid expansion, known as inflation. 
It can now solve several cosmological problems, like the horizon, flatness and the magnetic-monopole problems of the present Universe.
In the SM, there are no more candidates to propose as candidates of dark matter and inflation, explaining neutrino variables or baryon asymmetry of the Universe (BAU). All these pieces of evidence suggest adding new degrees of freedom to the SM.

It is well known that there is an asymmetry in matter number density and the comoving baryon number density is small yet a non-zero quantity, $Y_B=(8.75\pm0.23)\times 10^{-11}$. The proper explanation of such observed non-zero baryon asymmetry of our Universe is beyond the reach of the SM framework. Interestingly, the very existence of neutrino mass has established the leptogenesis mechanism as a suitable choice to explain the asymmetry in the baryon number density of our Universe. Seesaw mechanisms \cite{Minkowski:1977sc, Yanagida:1979as, Gell-Mann:1979vob, Mohapatra:1979ia} are the dominant frameworks to explain neutrino mass in a more straightforward and precise way \cite{Hambye:2012fh, Fukugita:1986hr, Das:2018qyt}. In particular, the three fundamental conditions to create baryon asymmetry, known as the ``three Sakharov conditions" \cite{Sakharov:1967dj} are also well adequate with seesaw frameworks. In the leptogenesis mechanism, one looks for lepton number violation interactions, satisfying the ``three Sakharov conditions", which can produce sufficient lepton asymmetry that eventually converts to baryon asymmetry via the sphaleron process. Tree-level seesaw mechanisms are categorized into three types depending on the type of heavy mediators to generate neutrino mass. Baryogenesis via leptogenesis is widely studied in these all types of seesaw frameworks and readers are referred to \cite{Buchmuller:2004nz,Davidson:2008bu,DeSimone:2007edo,Frossard:2012pc} for type-I, \cite{Antusch:2004xy, Bazzocchi:2009da,Pascoli:2006ci,Strumia:2006qk,Barrie:2022cub} for type-II and \cite{Abada:2008ea, Mishra:2020gxg,Gu:2010xc,Adhikari:2010yt, Gopalakrishna:2018uxn, Franceschini:2008pz, Foot:1988aq, Schechter:1980gr, Zhang:2009ac, He:2012ub,Goswami:2018jar} for type-III seesaw references. 

In this work, we extend the SM 	by three hyperchargeless $Y=0$ triplet fermions, i.e., vector-like fermions $\Sigma_{i=1,2,3}$  and a Higgs triplet $\Delta$ with hypercharge $Y=2$ \cite{Das:2020uer,Goswami:2018jar}.
The first two generation fermion triplets $\Sigma_{1,2}$ are even under the $Z_2$ symmetry transformation, while the third one, $\Sigma_{3}$ and scalar triplet, $\Delta$ are odd under the same transformation.	
This framework unifies the simultaneous explanation of the neutrino mass and mixings, dark matter, baryogenesis via leptogenesis, inflation and reheating temperature of the Universe.
The $Z_2$ even neutral fermions from the first two triplets  $\Sigma_{1,2}$  explain the neutrino mass and mixing angles. The $Z_2$ even fermion triplets mix among themselves and decay processes involving the lepton and Higgs can explain the observed BAU value via the resonant leptogenesis \cite{Pilaftsis:2003gt} process. Meanwhile, the third triplet fermion, $\Sigma_{3}$, serves as a viable dark matter candidate and can saturate the current relic density of the Universe. The scalar triplet $\Delta$ couples to gravity non-minimally, the real part of the neutral component, that is, heavy Higgs, can act as inflation.
We obtain all the inflationary parameters and find them according to the new constraints from Planck-2018 and related experiments \cite{Planck:2018jri,BICEPKeck:2022mhb}. The light SM particles from the decays/annihilations of the inflaton after inflation can reheat the Universe again.
We also check that the interaction terms for both the $Z_2$ even and odd sectors can provide the observed baryon asymmetry of the Universe at the TeV scale.

The rest of the paper is organized as follows. We present the details of the model in section~\ref{sec:model}. The masses and relevant interaction couplings for all the particles are also presented in section~\ref{sec:model}.
Afterwards, we introduce the theoretical and experimental constraints in section~\ref{sec:bounds}. 
The possible explanation for neutrino low energy variables, dark matter, leptogenesis and inflation and reheating have been discussed in sections.~\ref{sec:numass}-\ref{sec:lptgns}. We then conclude our analysis in section~\ref{sec:concl}.

\section{Model Framework}
\label{sec:model}
This section will provide a detailed discussion of our workable model.
The SM-like Higgs doublet and other additional model particles in this present framework are shown in table~\ref{tmodel}.
\begin{table}[h!]
	\centering
	\begin{tabular}{|c|c|c|c|c|c|}
		\hline
		Fields&Representation   & $SU(3)$&$SU(2)$&$Y$&$Z_2$ \\
		\hline
		$\Phi$&$\begin{pmatrix}
			\phi^\pm\\ 1/\sqrt{2}(v+h)\\
		\end{pmatrix}$&1&2&$\frac{1}{2}$&$+1$\\
		$\Sigma_i$&$\begin{pmatrix}\Sigma^0/\sqrt{2}&\Sigma^+\\\Sigma^-&-\Sigma^0/\sqrt{2}
		\end{pmatrix}_{i=1,2}$&1&3&0&$+1$\\
		$\Sigma_3$&$\begin{pmatrix}\Sigma^0/\sqrt{2}&\Sigma^+\\\Sigma^-&-\Sigma^0/\sqrt{2}
		\end{pmatrix}_{3}$&1&3&0&$-1$\\
		$\Delta$&$\begin{pmatrix} H^+/\sqrt{2}&H^{++}\\\frac{H+i A}{\sqrt{2}}&-H^+/\sqrt{2}\end{pmatrix}$&1&3&2&$-1$\\
		\hline
	\end{tabular}
	\caption{Fields and their corresponding charge content}
	\label{tmodel}
\end{table}

The renormalizable Lagrangian for triplet fermions is given as,
\bea
\mathcal{L}= &&Tr[\overline{\Sigma_i}i\slashed{D}\Sigma_i]-\frac{1}{2}Tr[\overline{\Sigma_i}M_{\Sigma_i}\Sigma_i^c]-Y_{\Sigma_j}\big( \Tilde{\Phi}^\dagger\overline{\Sigma_j}L\big)+\frac{1}{2}Y_{3i}\big(Tr[\overline{\Sigma_3}\Delta] \ell_{Ri}\big)+h.c. \label{lag1},
\eea
where $i=1,2,3$ and $j=1,2$ in the subscript represent generation indices for the new fermion triplets. The $j=1,2$ stand for the first two triplet fermions, even under $Z_2$ transformation. The notation $j=3$ breaks the $Z_2$ symmetry; hence we do not have it in the Lagrangian. The first two terms represent the kinetic and mass terms of the triplet fermions. The third term in the Lagrangian involves with neutrino low energy variables and baryogenesis. In contrast, the last term is associated with dark matter and inflation depending on the choice of parameters. The last term of the Lagrangian (equation~\eqref{lag1}) can be further expanded as follows,
\be
\frac{1}{2}Y_{3i}\, Tr[\overline{\Sigma_3}\Delta] \ell_{Ri} =\frac{1}{2}Y_{3i} \, \Big(\Sigma_3^-H^{++}+\bar{\Sigma_3^0}H^++\Sigma_3^+\frac{\left(H+i A\right)}{\sqrt{2}} \Big) \ell_{Ri}.
\label{eq:mixterD}
\ee
{The triplet scalar $\Delta$ is $Z_2$ odd in this model and it does not acquire any vacuum expectation value (VEV)\footnote{For large values of the Yukawa couplings $Y_{3i}$, the existing $Z_2$ may break and neutral scalar fields of the triplet may acquire non-zero VEV(s) at a high scale depending on the other parameters. It needs a detailed analysis which is out of the scope of the present work. We use small Yukawa, including large scalar quartic couplings, to avoid such scenarios.}. However, the Higgs doublet does acquire a non-zero VEV. After electroweak symmetry breaking and the scalar potential can be expressed as follows},
\bea
V(\Phi,\Delta)=&&-m_\Phi^2 |\Phi|^2+m_\Delta^2Tr[\Delta^\dagger\Delta]+\frac{\lambda_1}{4}(\Phi^\dagger\Phi)^2+\lambda_2[Tr(\Delta^\dagger\Delta)]^2\nonumber\\&&+\lambda_3Tr[(\Delta^\dagger\Delta)^2]+\lambda_4\big(\Phi^\dagger\Delta\Delta^\dagger\Phi\big)+\lambda_5(\Phi^\dagger\Phi)Tr(\Delta^\dagger\Delta). \label{pot1}
\eea
After the electroweak symmetry breaking of the scalar potential~\eqref{pot1}, we get seven massive physical eigenstates ($H^{\pm\pm}, H^\pm, A, H$ and $h$). There are three unphysical massless eigenstates, i.e., the three Goldstone bosons $G^\pm, G^0$, which are eaten up to give mass to the gauge bosons $W^\pm$ and $Z$. The masses of the physical scalars at the tree level can be expressed as follows,
\begin{eqnarray}
	M_h^2=&&2\lambda_1v^2,\\
	M^2_H=&& m_\Delta^2-\frac{1}{2} (\lambda_4+\lambda_5) v^2,\\
	M^2_A=&& m_\Delta^2-\frac{1}{2} (\lambda_4+\lambda_5) v^2,\\
	M^2_{H^{\pm}}=&&m_\Delta^2-\frac{1}{4} (\lambda_4+2 \lambda_5) v^2,\\
	M^2_{H^{\pm\pm}}=&&m_\Delta^2-\frac{\lambda_5v^2}{2}.
\end{eqnarray}
It is to be noted that both the masses of CP-even $H$ and CP-odd (pseudoscalar) $A$, degenerate at the tree level. Either of them could explain the inflation parameters. We will discuss it later in detail. 
\section{Bounds on the models}
\label{sec:bounds}
Theoretical considerations like absolute vacuum stability, perturbativity, and unitarity of the scattering matrix constrain the parameter space of this model. The following will discuss these theoretical bounds and the bounds from the electroweak precision measurements on the present model parameters.

\subsection{Constraints from the stability of scalar potential}
The stability of the electroweak vacuum of the scalar potential in equation~\eqref{pot1} requires that it should be bounded from below, i.e., there is no direction in field space along which the potential tends to minus infinity. The conditions are~\cite{Arhrib:2011uy,Moultaka:2020dmb}
\allowdisplaybreaks
\bea
&&\lambda_1 (\Lambda) \geq 0,~~\lambda_2(\Lambda)+\lambda_3(\Lambda)\geq 0, ~~\lambda_2(\Lambda)+\frac{\lambda_3(\Lambda)}{2} \geq 0,\nn \nn\\
&&\lambda_5(\Lambda)+ \sqrt{\lambda_1(\Lambda)(\lambda_2(\Lambda)+\lambda_3(\Lambda))}\geq 0,\nn\\
&&\lambda_5(\Lambda)+ \sqrt{\lambda_1(\Lambda)\left(\lambda_2(\Lambda)+\frac{\lambda_3(\Lambda)}{2}\right)} \geq 0,\nn\\
&&\lambda_5(\Lambda)+\lambda_4(\Lambda)+\sqrt{\lambda_1(\Lambda)(\lambda_2(\Lambda)+\lambda_3(\Lambda))}\geq 0,\nn\\ &&{\rm and}~~\lambda_5(\Lambda)+\lambda_4(\Lambda)+ \sqrt{\lambda_1(\Lambda)\left(\lambda_2(\Lambda)+\frac{\lambda_3(\Lambda)}{2}\right)} \geq 0.\nn
\label{eq:lstab}
\eea
where the coupling constants are evaluated at running scale $\Lambda$. {We presented the corresponding Renormalization Group Equations (RGEs) for this model in the Appendix~\ref{sec:app}. We used the proper matching conditions at different scales up to top mass $M_t$, added the new physics part at a new scale (consider dark matter mass scale), and then ran up to $\Lambda$.}
\subsection{Perturbativity and Constraints from unitarity of the scattering matrix}
For the model to behave as a perturbative quantum field theory at any given scale, one must impose the conditions on the radiatively improved scalar potential $V(\Phi,\Delta)$ as, 
\allowdisplaybreaks
\beq
\mid \lambda_{1,2,3,4,5}(\Lambda)\mid \leq 4 \pi.
\label{pertHT2}\nn
\eeq
The tree-level unitarity of the S-matrix for elastic scattering imposes the following constraints~\cite{Arhrib:2011uy},
\allowdisplaybreaks
\begin{eqnarray}
	&&|\lambda_5(\Lambda)+\lambda_4(\Lambda)| \leq 8\pi,~~
	|\lambda_5(\Lambda)| \leq 8\pi\,,~~
	|2\lambda_5(\Lambda)+3\lambda_4(\Lambda)|\leq 16\pi,\nn\\
	&&|\lambda_1(\Lambda)| \leq 16\pi\,\nonumber,~~
	|\lambda_2(\Lambda)| \leq 4(\Lambda)\pi\,~~
	|\lambda_2(\Lambda)+\lambda_3(\Lambda)| \leq 4\pi,\\
	&& \Big |\lambda_1(\Lambda)+4\lambda_2(\Lambda)+8\lambda_3(\Lambda)\pm\sqrt{(\lambda_1(\Lambda)-4\lambda_2(\Lambda)-8\lambda_3(\Lambda))^2+16\lambda^{2}_4}(\Lambda)\Big |\leq 32\pi, \nn\\
	&&\Big |3\lambda_1(\Lambda)+16\lambda_2(\Lambda)+12\lambda_3(\Lambda)
	\pm\sqrt{(3\lambda_1(\Lambda)-16\lambda_2(\Lambda)-
		12\lambda_3(\Lambda))^2+24(2\lambda_5(\Lambda)
		+\lambda_4(\Lambda))^2} \Big |\leq 32\pi,\nonumber \\
	&&|2\lambda_5(\Lambda)-\lambda_4(\Lambda)|\leq 16\pi\,~~{\rm and}~~|2\lambda_2(\Lambda)-\lambda_3(\Lambda)|\leq 8\pi.\nonumber
	\label{eq:luni}
\end{eqnarray}
These conditions imply an upper bound on the couplings $\lambda's$ at an energy scale $\Lambda$. 
\subsection{Constraints from the electroweak precision experiments}\label{prey2}
At the loop-level, the contributions of the scalar triplet with hypercharge $Y=2$ to the $S$, $T$ and $U$ parameters are given by~\cite{Lavoura:1993nq, Chun:2013vca}, 
\allowdisplaybreaks
\begin{eqnarray}
	S^{Y=2} &=& -\frac{1}{3\pi} \ln\frac{m_{+1}^2}{ m_{-1}^2}
	-\frac{2}{\pi} \sum_{T_3=-1}^{+1} (T_3 - Q s_W^2)^2 \,
	\xi\left(\frac{m_{T_3}^2}{ m_Z^2}, \frac{m_{T_3}^2}{ m_Z^2}\right), \nn\\
	T^{Y=2} &=& \frac{1}{ 16\pi c_W^2 s_W^2} \sum_{T_3=-1}^{+1} \left(2-T_3(T_3-1)\right)\,
	F\left(\frac{m_{T_3}^2}{ m_Z^2}, \frac{m_{T_3-1}^2}{ m_Z^2}\right), \nonumber\\
	U^{Y=2} &=& \frac{1}{6\pi} \ln \frac{m_{0}^4 }{ m_{+1}^2 m_{-1}^2}
	+\frac{1}{\pi} \sum_{T_3=-1}^{+1} \left[ 2(T_3 - Q s_W^2)^2\,
	\xi\left(\frac{m_{T_3}^2}{ m_Z^2}, \frac{m_{T_3}^2}{ m_Z^2}\right) \right.\nonumber\\
	&& \left. ~~~~~~~~~~~~~~~~~~~~~~~~~~~~
	-(2-T_3(T_3-1))\, \xi\left(\frac{m_{T_3}^2}{ m_W^2}, \frac{m_{T_3}^2}{ m_W^2}\right) \right], \nonumber
\end{eqnarray}
where $m_{+1,0,-1} \equiv M_{H^{++},H^+,H}$ and the function $\xi(x,y)$ is defined as~\cite{Lavoura:1993nq},
\allowdisplaybreaks
\begin{eqnarray}
	\xi (x_1, x_2) & = &
	\frac{4}{9} - \frac{5}{12} (x_1 + x_2) + \frac{1}{6} (x_1-x_2)^2
	\nonumber\\
	&  &
	+ \frac{1}{4} \left[ x_1^2 - x_2^2 - \frac{1}{3} (x_1-x_2)^3 -
	\frac{x_1^2 + x_2^2}{x_1 - x_2} \right] \ln \frac{x_1}{x_2}
	\nonumber\\
	&  &
	- \frac{1}{12} \Delta(x_1, x_2) f(x_1, x_2)\, .
	\label{eq:csi}\nn
\end{eqnarray}
The definitions of $\Delta$, $f$ and $F$ can be written as

\bea 
f(x_1,x_2)&=&\left\{
\begin{array}{ll}
	-2\sqrt{\Delta}\big\{ \arctan\frac{x_1-x_2+1}{\sqrt{\Delta}}
	-\arctan\frac{x_1-x_2-1}{\sqrt{\Delta}}\big\}\,,
	&(\Delta>0)\,  \\ 
	0\,,
	&(\Delta=0)\,\\\label{fxy}
	\sqrt{-\Delta}\ln\frac{x_1+x_2-1+\sqrt{-\Delta}}{x_1+x_2-1-\sqrt{-\Delta}}\,,
	&(\Delta<0)\,,
\end{array}
\right.
\eea
with $\Delta=2(x_1+x_2)-(x_1-x_2)^2-1$, $x_i\equiv m_i^2/q^2$, {where $q$ being the arbitrary mass parameter used in dimensional regularization and,}
\bea
F(m_1^2,m_2^2) &=& F(m_2^2,m_1^2) 
= \frac{m_1^2+m_2^2}2 -\frac{m_1^2 m_2^2}{m_1^2-m_2^2}\ln \left(\frac{m_1^2}{m_2^2}\right). \label{Fm1m2}   
\eea

These parameters can constrain the model parameter space from the electroweak precision data. From the recent precision data the oblique parameters are measured as $S = −0.01\pm 0.07 \text{ and } T = 0.04\pm  0.06$~\cite{ParticleDataGroup:2022pth}. The new triplet mass heavily suppresses the $U$ parameter and can be considered zero compared to the $S$ and $T$ parameters.
\subsection{Constraints from the LHC}

The direct search of $H^{\pm\pm}$ via $pp \ra H^{++} H^{--}, ~H^{\pm\pm}\ra W^{*\pm} W^{*\pm}\ra$ $ \mu^{\pm} \nu_\mu \mu^{\pm}\nu_\mu$ process at the LHC, puts a lower bound on $M_{H^{\pm\pm}}>84$ GeV~\cite{Kanemura:2014ipa,Primulando:2019evb,Ashanujjaman:2021txz,Mandal:2022ysp,Dev:2021axj,Dziewit:2021pak,Bai:2021ony,Dey:2020tfq,Fuks:2019clu}. The recent direct LHC search ($\sqrt{s}=13$ TeV with integrated luminosity $\mathcal{L}=139~{\rm fb}^{-1}$) of $H^{\pm\pm}$ via $pp \rightarrow H^{++}H^{--}, H^{\pm\pm}\rightarrow W^{\pm\pm}W^{\pm\pm}$ (on-shell) process at the LHC excludes 200 GeV$<m_{H^{\pm\pm}}<350$ GeV~\cite{Ashanujjaman:2021txz}. If constraints like the stability, unitarity, $T$-parameter and $\mu_{\gamma\gamma}$ at LHC are considered,  then one can obtain the following lower bounds on the non-standard scalar masses: $M_{H^+}>130$ GeV, $M_{A,H}>150$ GeV~\cite{Das:2016bir}.

\section{Neutrino mass}
\label{sec:numass}

Recalling the Lagrangian from equation.~\eqref{lag1}, the terms responsible for neutrino mass generation are \cite{Goswami:2018jar, Foot:1988aq, Schechter:1980gr,Zhang:2009ac,He:2012ub}:
\be
\mathcal{L}= -\frac{1}{2}Tr[\overline{\Sigma_i}M_{\Sigma_i}\Sigma_i^c]-Y_{\Sigma_j}\big( \Tilde{\phi}^\dagger\overline{\Sigma_j}L\big).
\ee
Since the third generation, $\Sigma_3$ of the fermion triplet is $Z_2$ odd, only $\Sigma_{1,2}$ will participate in the neutrino mass generation process. We have considered degenerate heavy-fermion masses ($M_\Sigma=M_{\Sigma_1}=M_{\Sigma_2}$), so one can consider the Majorana mass matrix $M$ is proportional to the identity matrix.
After the electroweak symmetry breaking, the neutrino mass matrix takes the form,
\be
M_\nu=\begin{pmatrix}0&M_D^T\\M_D&M_\Sigma\end{pmatrix}.
\ee
Here, $M_D=Y_\Sigma \, v/\sqrt{2}$ and $v=246.221$ GeV is the VEV of the doublet Higgs. The given mass matrix can be diagonalized by a unitary matrix (say $U_f$)\cite{Goswami:2018jar}  with $U_f^TM_\nu U_f=m_{Diag}$. The diagonalized mass matrix consists of three light neutrino mass eigenvalues (the lightest one being zero in this case) and two heavy Majorana mass eigenvalues, eventually taking the form of $(0,m_2,m_3,M,M)$. The effective light neutrino mass matrix can be expressed as,
\be
m_\nu^{eff}=-(v^2/2)Y_\Sigma.M_\Sigma^{-1}.Y_{\Sigma}^T.
\ee
In our scenario, the Yukawa matrix is a ($3\times2)$ matrix in the flavour space due to the two triplet generations involved in the visible sector; hence it contains new sources of CP violation. We can parametrize the $Y_\Sigma$ matrix employing the well-known Casas-Ibarra parametrization~\cite{Casas:2001sr, Ibarra:2003up} to ensure the exact low energy parameters as,
\be
Y_\Sigma=\frac{\sqrt{2}}{v}U_{PMNS}^*.\sqrt{\hat{m_\nu}}.R.\sqrt{\hat{M_\Sigma}},
\ee
here, $U_{PMNS}$ is the Pontecorvo-Maki-Nakagawa-Sakata (PMNS) mixing matrix, which diagonalizes the effective neutrino mass $m_\nu^{eff}$. $\hat{m_\nu}$ and $\hat{M_\Sigma}$ are the diagonal matrix of the square roots of the eigenvalues of the $m_{\nu}^{eff}$ and $M_{\Sigma}$ respectively and $R$ is an orthogonal complex matrix, which can be expressed as~\cite{Goswami:2018jar}
\be
R=\begin{pmatrix}0&0\\
	cos z& \sin z\\
	-\chi\sin z&\chi \cos z\end{pmatrix}.
\ee
For two triplet generations, we only have a single complex parameter $z$, expressed as $z=x+iy$, with $x,y\in[0,2\pi]$ \cite{Ibarra:2003up} for each complex plane. We fix the value of $\chi=1$ for our entire analysis. The light neutrino masses are determined by the entries of the $R$ matrix and the low energy observables associated with the unitary matrix $U_{PMNS}$. In such cases, the light triplet ($\mathcal{O}$(TeV)) does not necessarily imply small values of neutrino masses.

\begin{figure}[h!]
	\begin{center}
			\includegraphics[scale=0.85]{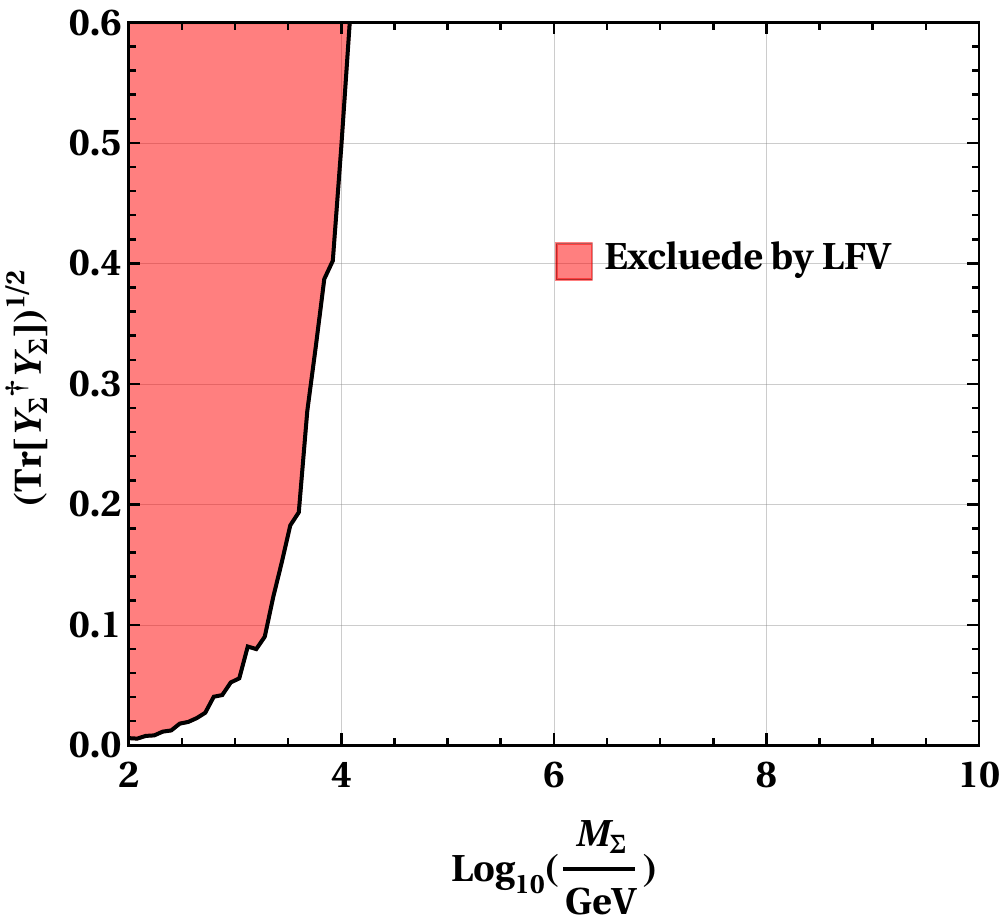}
				\caption{\label{fig:neu} The shaded region is excluded from the flavour violating decay $(\mu \to e \gamma)$ process~\cite{SINDRUMII:1993gxf,Abada:2008ea}. However, the whole parameter space can satisfy light neutrino mass and mixing angle via Casas-Ibarra parametrization~\cite{Goswami:2018jar}.} 
			\end{center}
		\end{figure}

We have shown the plots in Fig.~\ref{fig:neu} which (all the regions) explain the neutrino mass using Casas-Ibarra parametrization \cite{Casas:2001sr}. But the flavor violating decay process specially $BR(\mu \to e \gamma)$ and $BR(\tau \to eee)$ put a stringent bound on it~\cite{SINDRUMII:1993gxf,Abada:2008ea}. 
Interestingly, the Yukawa couplings in equation~\eqref{eq:mixterD} can also influence and modify these branching ratios along with the muon as well as electron anomalous magnetic moments. However, in this model, we avoided such situations as the contributions to these decay processes and both the anomalous magnetic moments~\cite{Abi:2021gix,Chakrabarty:2022voz} are negligible due to large fermionic and scalar masses. In the parameter region on which is focused in this manuscript, the branching fractions of the LFV processes are much smaller than the current bounds.

		\section{Dark matter}
		\label{sec:DM}
		The viable DM candidate in this model is the lightest $Z_2$-odd singlet scalar $H, A$ or the fermion $\Sigma_3^0$. Here, the relic abundance of the DM candidate has been achieved through the Freeze-out and/or Freeze-in mechanism, depending on the choice of parameter spaces. Suppose the dark matter is in thermal equilibrium in the early Universe, then $T>M_{DM}$, where $T$ is the temperature of the Universe, and it Freezes out when $T<M_{DM}$. If it is not in thermal equilibrium in the early Universe, in that case, it could have been produced from some mother (heavy) particles and given the correct relic density through the Freeze-in mechanism.
		
		Dark matter, produced from the decay or annihilation of various mother particles, is in thermal equilibrium in the early Universe only when the interaction rate is greater than the Hubble expansion rate. This condition can be written as, 
		\beq
		{\Gamma \over H(T) } \geq 1,
		\eeq
		where, $\Gamma$ is the relevant interaction rate and  $H(T)$ is the Hubble parameter given by~\cite{Plehn:2017fdg,Hall:2009bx},
		\beq
		H(T) = \left( g^* \, \frac{\pi^2}{90} \, \frac{T^4}{\mpl^2} \right)^{1/2},
		\label{eq:Hub}
		\eeq
		where, $\mpl=2.4\times 10^{18}$ GeV is the reduced Planck mass.
		If the production of mother particles occurs mainly from the annihilation of other particles in the thermal bath, the interaction rate, $\Gamma$ will be replaced by~\cite{Plehn:2017fdg,Hall:2009bx},
		\beq
		\Gamma = n_{\rm eq} <\sigma v>,
		\eeq
		where, $n_{\rm eq}$ is their equilibrium number density and is given by~\cite{Plehn:2017fdg}
		\beq
		\begin{aligned}
			&n_{\rm eq} &=
			&\left\{ \begin{array}{l} \vspace{0.3cm} g^* \left( \frac{m T}{2 \pi}\right)^{3/2} \, e^{-m/T},  ~~~~~~~{\rm for ~non\text{-}relativistic ~states}~~T<<M\\
				
				\frac{\zeta_3}{\pi^2} g^* T^3,  ~~~~~~~~~~~~~~~~~~~~{\rm for~relativistic~boson ~states}~~T>>M\\
				
				\vspace{0.5cm}
				\frac{ 3}{4}\, \frac{\zeta_3}{\pi^2} g^* T^3 ,  ~~~~~~~~~~~~~~~~~~{\rm for~relativistic~fermion ~states}~~T>>M
				\vspace{-0.2cm}\end{array}
			\right.
			\label{eq:n}
		\end{aligned}
		\eeq
		where, the Riemann zeta function has the value $\zeta_3=1.2$ and $g^*$ is the effective degree of freedom in this framework. Here,
		$<\sigma v>$ is the thermally averaged annihilation cross-section of the particles in the thermal bath and can be expressed as~\cite{Gondolo:1990dk,Plehn:2017fdg},
		\beq
		<\sigma_{xx} v> = \frac{  2 \pi^2 T \, \int_{4 m^2}^\infty ds \sqrt{s} \, (s-4 m^2)  \, K_1(\frac{\sqrt{s}}{T})  \sigma_{xx}   }{   \left( 4 \pi m^2 T K_2(\frac{m}{T})  \right)^2        },
		\eeq
		where, {$K_{1(2)}$ is the modified Bessel function of the first (second) kind}. The dark matter is in thermal equilibrium at early Universe, $i.e.$, $\frac{ n_{eq}<\sigma_{xx} v>}{H(T)} >> 1$.
		In this work, we find that the non-thermally produced dark matter can not serve as a viable dark matter candidate due to the large production rate, while it can produce exact relic density through the Freeze-out mechanism, which we will discuss now.
		
		The lightest one between $H$ and $A$ can also serve as a viable WIMP dark matter candidate, which may saturate the measured DM relic density of the Universe at the current time. In this model, the dark matter candidate $H$ or $A$
		can annihilate to the SM particles via a Higgs (125 GeV) exchange or a $Z$ boson through $s$-channel diagrams and $H$, $A$ and $H^\pm$, $H^\mp$ mediated $t$- and $u$-channel diagrams. As the $H$ and $A$ can interact with the nucleons through the Higgs (125 GeV) and $Z$ mediated $t$-channel exchanges, the dark matter direct detection cross-sections are relatively large in this model~\cite{Araki:2011hm}. Hence, all the regions reached by direct detection experiments are ruled out here.
		
\begin{figure}[h!]
			\begin{center}
					\includegraphics[scale=0.24]{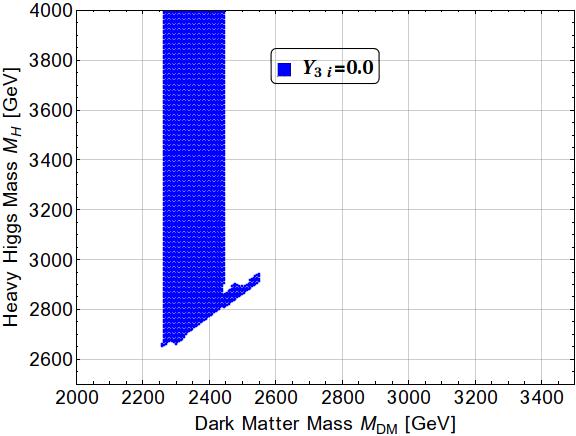} \hspace{0.2cm}
					\includegraphics[scale=0.24]{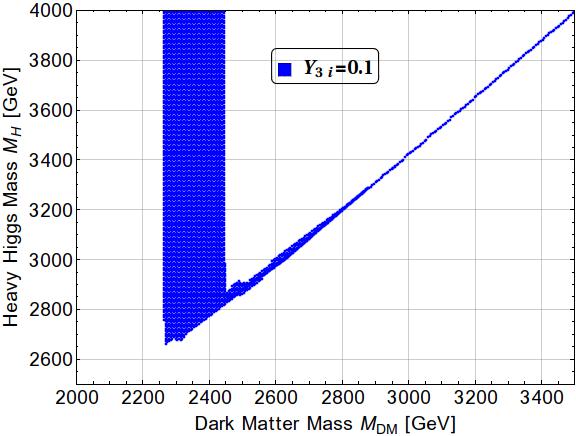} \hspace{0.2cm}
					\includegraphics[scale=0.24]{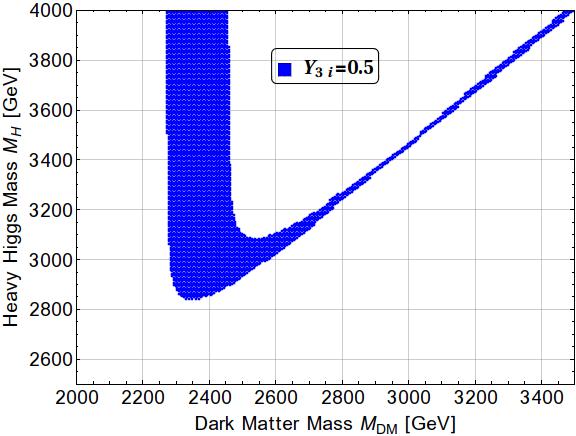}
					\caption{\label{fig:DM}{Dark Matter mass $vs.$ heavy scalar mass for three different values of  $Y_{3i}=0$, $0.1$ and $0.5$, respectively. The Heavy Higgs masses are $M_{A}=M_{H}$, $M_{H^\pm}=M_{H}+50$ GeV, $M_{H^{\pm\pm}}=M_{H}+100$ GeV. The blue band gives the exact relic density $\Omega h^2=0.1198\pm 0.0012$ within $3\sigma$.}}
				\end{center}
			\end{figure}

		The neutral $Z_2$-odd fermion $\Sigma_3^0$ could be a viable WIMP dark matter candidate, providing DM relic density depending on the model paramters. In this case, we get the exact relic density for the dark matter mass region greater than $2$ TeV. At tree level, the mass of the neutral $\Sigma_3^0$ and the charged fermions $\Sigma_3^\pm$ are degenerate. When considering the radiative one-loop correction, the charged fermions become slightly heavier than the neutral ones. The mass difference between them is given by~\cite{Cirelli:2009uv,Cirelli:2005uq},
		\beq
		\Delta M=(M_{\Sigma_3^\pm}-M_{\Sigma_3^0})_{1\text{-loop}}=\frac{\alpha M_{\Sigma_3}}{4\pi}\Big[f\Big(\frac{M_W}{M_{\Sigma_3}}\Big) -c_W^2 f\Big(\frac{M_Z}{M_{\Sigma_3}}\Big)\Big]
		\label{massdifftrip},
		\eeq
		with,
		$f(x)=-\frac{x}{4}\Big\{ 2 x^3 ~ {\rm log}(x)+(x^2-4)^{\frac{3}{2}}~ {\rm log}\left( \frac{x^2-2-x\sqrt{x^2-4}}{2} \right)\Big\}$. Authors of Refs~\cite{Cirelli:2009uv,Cirelli:2005uq} have shown that the mass splitting between charged and neutral fermions remains $\sim 160$ MeV for $M_{\Sigma_3} = 0.05-5$ TeV. As $\Delta M$ is very small, the effective annihilation cross-section is always dominated by the co-annihilation channels $\Sigma_3^0 \Sigma_3^\pm, \Sigma_3^\pm \Sigma_3^\pm \ra { SM ~particles}$~\cite{Griest:1990kh}. The effective annihilation cross-section becomes very large for the low dark matter mass region and DM production gets under abundance. However, this region is not ruled out, but we need some other component to get the exact relic density. In this model, we find that dark matter mass range ($2.285<M_{DM}<2.445$) TeV always provides exact relic density.		
		In contrast, we also checked that if the scalar triplet masses become closer to the $Z_2$ odd neutral fermion masses, another co-annihilation channel changes the dark matter parameter space. We present these co-annihilation effects to the relic density in the $M_H-M_{DM}$ plane in Fig.~\ref{fig:DM}.
		We keep the heavy Higgs masses at $M_{A}=M_{H}$, $M_{H^\pm}=M_{H}+50$ GeV, $M_{H^{\pm\pm}}=M_{H}+100$ GeV. The blue band gives the exact relic density $\Omega h^2=0.1198\pm 0.0012$ within $3\sigma$.
		The band near ($2.285<M_{DM}<2.445$) TeV is more crowded mainly due to the fermionic contribution. In the first plot, we use $Y_{3i}=0$; hence we have the dominant fermionic contribution. However, a small fraction of heavy Higgs contributes through the gauge and Higgs portal couplings for heavy Higgs masses close to the dark matter mass, i.e., the region below the blue band. The co-annihilation effect increases with the increase of Yukawa couplings $Y_{3i}$ ($i=1,2,3$), which can also be seen from rest of the plots in this Fig.~\ref{fig:DM}. It is clear from this figure that the presence of the Yukawa coupling can enhance the dark matter parameter space. 
		
			\section{Inflation and Reheating}
			\label{sec:infl}
			The present CMB data suggest the super-horizon anisotropies, measured by different experiments such as the Wilkinson Microwave Anisotropy Probe (WMAP), Planck. It is now proved that the early Universe underwent a period of rapid expansion known as inflation. This theory can explain several cosmological problems, such as the present Universe's flatness, horizon, and magnetic monopole problems.
			
			The experimental data suggest that the electroweak vacuum in the standard model is metastable. The Higgs quartic coupling remains negative at the GUT scale. Hence the Higgs is not a proper field to play the role of inflaton~\cite{Bezrukov:2007ep, Bezrukov:2008ut, Bezrukov:2008ej, Bezrukov:2009db,Okada:2015zfa,Barrie:2021mwi}. Therefore, we must need an extra new degree of freedom to explain the inflation of the Universe~\cite{Lerner:2009xg, Lebedev:2011aq}.
			
			Here, we study an extension of the SM Higgs sector with a complex triplet scalar $\Delta$ ($Y=2$) in the presence of large couplings $\zeta_{\phi,\Delta}$ to Ricci scalar curvature $R$ to explain inflation.

\begin{figure}[h!]
				\begin{center}
						\includegraphics[scale=0.3]{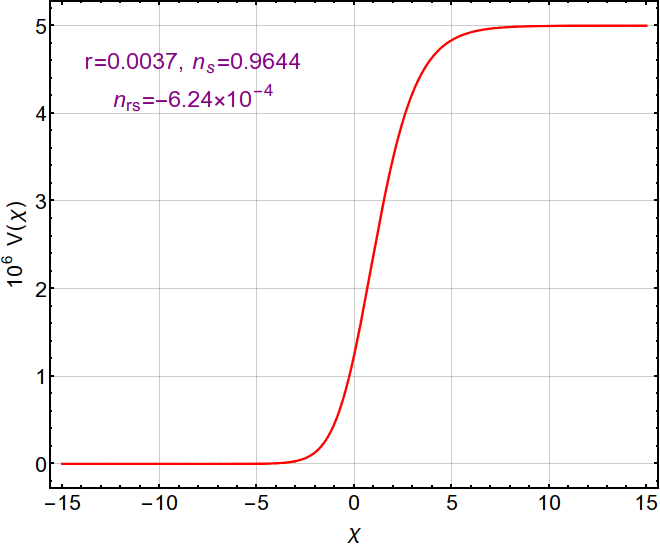} \hspace{0.2cm}
						\includegraphics[scale=0.314]{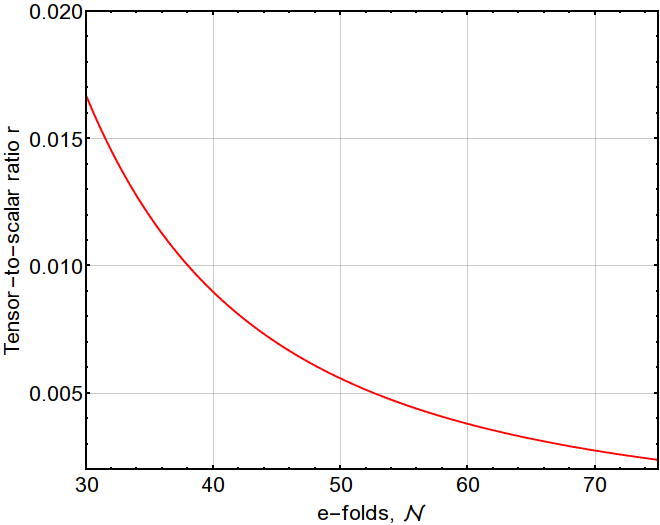}\\
						\includegraphics[scale=0.32]{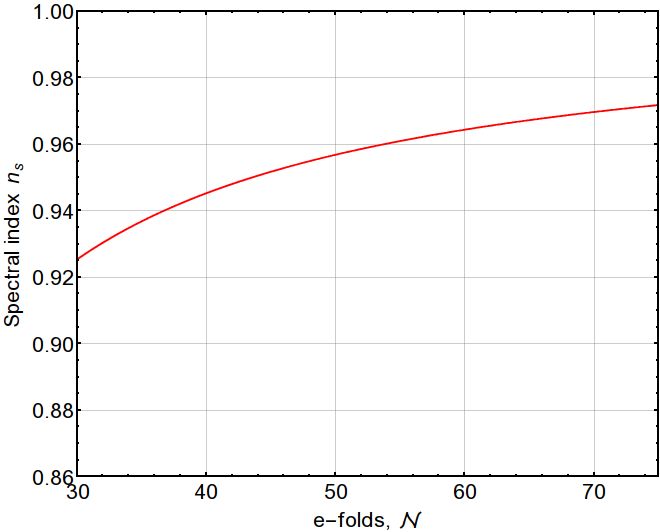}\hspace{0.2cm}
						\includegraphics[scale=0.320]{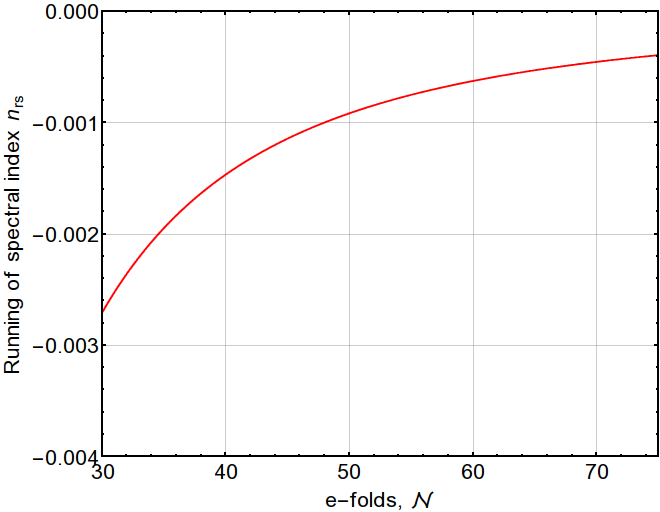}
						\caption{\label{fig:inflation} {Inflation potential in Planck unit, the tensor-to-scalar ratio $r$, the spectral index $n_s$ and the running of spectral index for $\zeta_\Delta=10^4$ and $\lambda_{2,3}=0.1$.} }
					\end{center}
				\end{figure}

\begin{figure}[h!]
					\begin{center}
							\includegraphics[scale=0.5]{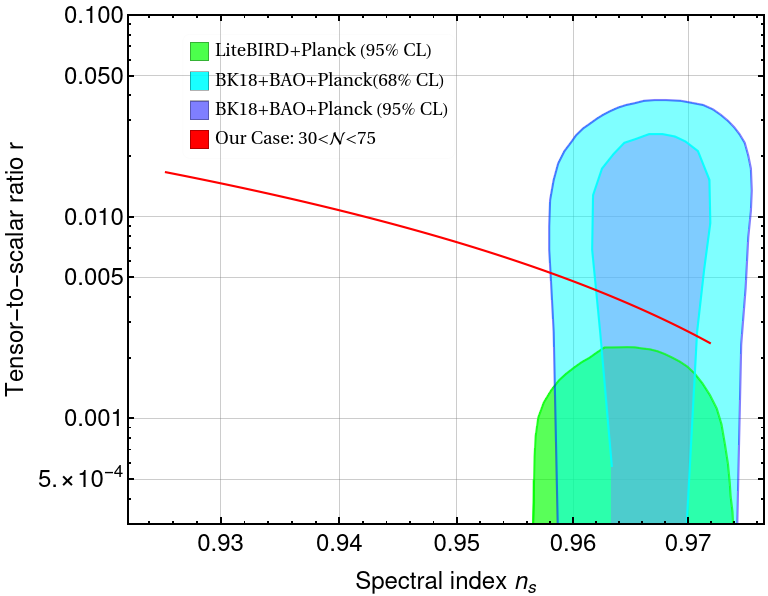} 
					\caption{Constrained regions in the $r	~vs.~ n_s$ plane in ``log" scale. The green region shows the upper limits from the $LiteBIRD+Planck$ \cite{LiteBIRD:2022cnt} data at 95\% CL. Cyan and violet regions are from BICEP/$Keck(18)+Planck$ \cite{BICEPKeck:2022mhb,BICEP:2021xfz, Planck:2018jri} data at 68\% and 95\% CL respectively. The red line stands $\mathcal{N}$ value within 30-75. }\label{fig:inflData}
						\end{center}
					\end{figure}

			The action of the fields in the Jordon frame is given by,
			\beq
			S_j = \int \sqrt{-g} d^4x\left[ {\cal L}_{SM} +  \frac{1}{2} (\partial_\mu \Phi)^\dagger (\partial_\mu \Phi)+\frac{1}{2}(\partial_\mu \Delta)^\dagger (\partial_\mu \Delta) - \zeta_\phi R |\Phi|^2 - \zeta_\Delta R |\Delta|^2 -V(\Phi,\Delta)\right].
			\label{Jorda}
			\eeq
			In this present work, we have interest for the inflation purely along $H$-direction\footnote{In this model, the Higgs $h$ and $A$ can also act as inflation in stable EW vacuum region.}, i.e., $h=0$, $A=0$, $H^\pm=0$ and $H^{\pm\pm}=0$. To calculate the inflationary observables such as spectral index $n_s$, tensor-to-scalar ratio $r$, etc., we perform a conformal transformation to the Einstein frame, where the non-minimal coupling $\zeta_\Delta$ of the scalar field to Ricci scalar disappears.
			The transformations is given by \cite{Kahlhoefer:2015jma},
			\beq
			\tilde{g}_{\mu\nu} = \Omega^2 g_{\mu\nu}, \quad\text{with}~~~\Omega=\sqrt{1+\zeta_\Delta \frac{H^2}{\mpl^2}}.
			\eeq
			The action of equation~\eqref{Jorda} in Einstein frame can be written as,
			\beq
			S = \int \sqrt{-g} \, d^4x\left[ \frac{1}{2}(\partial_\mu \chi)^\dagger (\partial_\mu \chi) -V(\chi)\right];
			\text{	with,}\quad
			\frac{d\chi}{dH} = \sqrt{\frac{\Omega^2 \mpl^2+ 6 \zeta_\Delta^2 H^2}{\Omega^4 \mpl^2}}.
			\eeq
			The scalar potential $V(\chi)$ is then given by,
			\beq
			V(\chi) = (\lambda_2 +\lambda_3) \frac{\mpl^4}{4 \zeta_\Delta^2} \left( 1+ exp\left( -\sqrt{\frac{2 \chi}{3 \mpl}}\right)\right)^{-2}.
			\label{eq:inflapotchi}
			\eeq
			The variation of scalar potential with $\chi$ are shown in Fig.~\ref{fig:inflation} (upper-left) for $\zeta_\Delta=10^4$ and $\lambda_{2,3}=0.1$ in the Planck unit. One can see that this potential shows a flat nature and supports slow-roll inflation.
				Using this scalar potential $V(\chi)$, one can define the slow-roll parameters $\epsilon,\eta~ \text{and}~ \zeta$ in terms of the potential as,
				\beq
				\epsilon =  \frac{1}{2} \left( \frac{1}{V} \frac{dV}{d\chi} \right),~~~
				\eta = \frac{1}{V} \frac{d^2V}{d\chi^2},~~~\text{and}~~~
				\zeta = \frac{1}{V^2} \frac{dV}{d\chi} \frac{d^3V}{d\chi^3} \nn.
				\eeq
				The inflationary observable quantities such as the tensor-to-scalar ratio $r$, the spectral index $n_s$ and the running of spectral index $n_{rs}$ are defined as,
				\beq
				r=16 \epsilon,~~~n_s = 1 -6 \epsilon + 2 \eta,~~~\text{and}~~~ n_{rs} = -2 \zeta -24 \epsilon ^2+16 \eta  \epsilon
				\eeq
				and the number of $e$-folds is given by,
				\beq
				\mathcal{N} = \int^{\chi_{end}}_{\chi_{start}}~ \frac{V}{dV/d\chi} d\chi
				\label{eq:efold}
				\eeq
				where $\chi_{start}$ ($\chi_{end}$) is the initial (final) value when the inflation starts (ends). We evaluate the integration assuming the slow roll parameter value starts from unity, $i.e.$, at $\chi_{start}$, $\epsilon=1$. We plotted the tensor-to-scalar ratio $r$, the spectral index $n_s$ and the running of spectral index for $\zeta_\Delta=10^4$ and $\lambda_{2,3}=0.1$ for different $e$-folds in Fig.~\ref{fig:inflation}. One can calculate $\chi_{end}$ from the above equation~\eqref{eq:efold} for $\mathcal{N}=60$. At the end of the inflation, we get the inflation observables as,
				\beq
				r=0.0037,~~ n_s=0.9644,~~~\text{and}~~ n_{rs}=- 6.24 \times 10^{-4},
				\eeq
				which are allowed by the present experimental data~\cite{Planck:2018jri,BICEPKeck:2022mhb,BICEP:2021xfz} as shown in Fig.~\ref{fig:inflData}. {Hence, the neutral component of the gauge triplet scalar can  serve as the inflaton in this model.}
					
					The energy density stored in the inflaton field, here $\chi$ ($\equiv H$) starts to disperse through the annihilation and/or decay into other particles, including those of the SM. This epoch is known as the reheating~\cite{Allahverdi:2010xz}. It takes the Universe from the matter-dominated phase during inflation to the radiation-domination phase. As $\chi$ falls below the Planck scale $\mpl$, the inflationary potential in equation~\eqref{eq:inflapotchi} can be approximated as a quadratic potential,
					\beq
					V_R=\frac{1 }{2} \omega^2 \chi^2,
					\eeq
					where, $\omega^2=\frac{(\lambda_2+\lambda_3) \mpl^2 }{3 \zeta_\Delta^2}$, which suggests the reheating occurs in this harmonic oscillator potential well as the field $\chi$ undergoes coherent oscillations with rapid frequency $\omega$~\cite{Linde:1981mu}. Now the equation of motion for inflation $\chi$ during reheating can be expressed as,
					\beq
					\frac{d^2\chi}{dt^2} + 3 H \frac{d\chi}{dt}+\frac{dV_R}{d\chi}=0,
					\label{eq:eqmreheat}
					\eeq
					here, $t$ stands for time, and $H$ is the Hubble expansion rate. In the limit $\omega >> H$, we get the solution from equation~\eqref{eq:eqmreheat} as,
					\beq
					\chi=\chi_0(t) \cos(\omega t),~\text{with}~\chi_0(t)=\sqrt{\frac{8 \zeta_\Delta^2}{(\lambda_2+\lambda_3)}}\, \frac{1}{t}.
					\eeq
					We now define $t_{\rm end}=\frac{2\zeta_\Delta}{\omega}$ as the time at which reheating stops and the amplitude become $\chi_{\rm end}=\sqrt{\frac{2}{3}}\frac{\mpl}{\zeta_\Delta}$. In this model, the inflaton, i.e., $\chi$ ($\equiv H$) can decay into the gauge bosons $W^\pm,Z$ through the kinetic coupling and pair of Higgs boson through quartic couplings $\frac{g^2}{4\sqrt{6}} \frac{\mpl}{\zeta_\Delta} \chi W^+ W^-$ and $\lambda_{4,5}\sqrt{\frac{2}{3}} \frac{\mpl}{\zeta_\Delta} \chi hh$ respectively. It is to be noted that the SM particles are massless at the time of reheating but get an effective mass due to the couplings to the inflaton and its oscillations. In the limit $\omega >> H$, one can write the masses at  constant $\chi_0$ as,
					\beq
					m_w^2=\frac{g^2}{2\sqrt{6}} \frac{\mpl}{\zeta_\Delta} |\chi|,~~~\text{and}~~~m_h^2=(\lambda_{4}+\lambda_{5})\sqrt{\frac{1}{6}} \frac{\mpl}{\zeta_\Delta} |\chi|.
					\eeq
					
					Hence, the effective coupling to the $W, Z$ bosons is large enough to be produced as a non-relativistic species. This is also true for the Higgs boson $h$ for $\lambda_{4,5}>1$. So the decay and annihilation of these gauges and Higgs bosons to the relativistic SM fermions will reheat the Universe.
					One can calculate the evaluation of number densities with the scale factor $a$ of the number density with the time for the gauge and Higgs bosons~\cite{Garcia-Bellido:2008ycs, Repond:2016sol, Borah:2018rca} as,
					\beq
					\frac{dn_i a^3}{dt}= A_i \omega a^3,
					\eeq
					where $i=w,h$. The {coefficients} in linear region are $A_w=\frac{P}{2 \pi^3}\, \frac{g^2 \mpl^2}{6 \zeta_\Delta^2} \sqrt{\frac{\lambda_2+\lambda_3}{2}} \chi_0(t_i)$ and $A_h=\frac{P}{2 \pi^3}\, \frac{(\lambda_4+\lambda_5) \mpl^2}{3 \zeta_\Delta^2} \sqrt{\frac{\lambda_2+\lambda_3}{2}} \chi_0(t_i)$  while at the resonance region $A_i=2Q \, n_i$ with $P=0.0455$ and $Q=0.045$.
					Here, $t_i$ is the time when the inflaton field becomes zero. The inflaton can decay into $W$ and Higgs bosons are only in the neighbourhood of $\chi=0$ when these bosons' masses (effective) are smaller than the inflaton mass $\omega$. It is to be noted that at low number densities of the produced bosons, their decays into SM fermions are always dominant for the production of relativistic particles and successful reheating of the Universe to the radiation-dominated epoch. However, the annihilation channels control the reheating temperature for large densities. 
					Parametric resonance production of $W$ bosons can occur only when the decay rate $\Gamma_W = \frac{3}{4} \alpha_W M_W$, falls below its resonance production rate at $\chi_0 = \frac{3.56}{\pi} \frac{Q^2 (\lambda_2+\lambda_3)}{\alpha_W^3} \chi_r\approx 60 (\lambda_2+\lambda_3) \chi_r $. It gives lower bounds $(\lambda_2+\lambda_3)>\frac{1}{60}$.
					
					As for the resonance production of Higgs bosons, this occurs when the decay rate of Higgs into fermions (governed by the Yukawa couplings $y_f$), $\Gamma=\frac{y_f^2 m_h}{16 \pi}$, falls below its resonance production rate for $\chi_0 \approx 0.41 \frac{(\lambda_2+\lambda_3)}{(\lambda_4+\lambda_5)} \chi_r $. We find that the gauge bosons will always dominate the other decay channels due to the large gauge couplings $g$, while the Higgs boson can be comparable with gauge bosons decay in the limit, $(\lambda_4+\lambda_5)\lesssim 0.006$~\cite{Borah:2018rca}. Even if we neglect the Higgs contributions to the entire energy density is $\rho_r=\frac{1.06\times 10^{57}}{\lambda_2+\lambda_3}$ GeV$^{4}$~\cite{Choubey:2017hsq,Borah:2018rca}. One can compute the reheating temperature as $T_r=\left( \frac{30 \rho_r}{\pi g_*} \right)^{\frac{1}{4}}=1.38\times 10^{14}$ GeV for $\lambda_{2,3}\approx 0.1$ and the number of degrees of freedom $g_*=139$ in the relativistic plasma that includes the SM particles plus new particles in this model. It is also to be noted that the contributions from the $\frac{1}{2}Y_{3i}\big(Tr[\overline{\Sigma_3}\Delta] \ell_{Ri}\big)$ term is neglected as compared to the gauge and Higgs bosons contributions.  
					
					\section{Leptogenesis with fermion triplets}
					\label{sec:lptgns}
					The triplet fermion leptogenesis is different from the conventional singlet fermion processes \cite{Pilaftsis:2003gt,Borah:2018rca,Das:2019ntw,Mahanta:2019gfe,Buchmuller:2004tu,Harz:2021psp} due to their gauge couplings \cite{Hambye:2012fh}.
					As the Universe expands, the triplet's mass exceeds the temperature of the Universe, and their equilibrium abundance gets Boltzmann suppressed. Two significant processes associate the triplets with their equilibrium abundance: annihilation into gauge bosons and decay into leptons and Higgs field. Typically the decay process is the CP-violating process and
					generates a net lepton asymmetry, and interestingly no asymmetry is generated in the annihilation process. The gauge coupling controls the annihilation, generally higher than the Yukawa coupling
					associated with decay. However, for temperatures below the triplet mass, the annihilation rate per triplet is Boltzmann suppressed since it is proportional to the number of
					triplets and at the same time to generate small neutrino masses, the Yukawa couplings controlling the decays and inverse decays are also small, hence
					the annihilation process is the dominant process. However, for comparatively larger values of neutrino masses, one can get larger values of the Yukawa couplings, and the decay dominates over annihilation, leading to an asymmetry until they get so large that decays and inverse decays are in thermal equilibrium, and any asymmetry gets washed out.
					In a short note of the triplet decay processes, at high temperatures, the gauge reactions are much faster than the expansion rate of the Universe; hence there is  no asymmetry produced at this stage. As the temperature drops, thermalization of the triplet distribution becomes less efficient
					and depending on the strength of the Yukawa interactions, the generation of the lepton asymmetry can proceed either after the decoupling of gauge reactions or after the Yukawa interactions freeze-out \cite{Hambye:2003rt}.
					
					In this work, we are studying leptogenesis scenarios where the lepton asymmetry is produced by the mass splitting of the $Z_2$-even fermion triplets $\Sigma_1$ and $\Sigma_2$ with {$\Delta m_{ij}=M_{\Sigma_i}-M_{\Sigma_j}$ ($i,j=1,2$)}.
					The most general form of CP asymmetry from the triplet decay can be expressed as \cite{Hambye:2003rt, Hambye:2012fh},
					\begin{eqnarray}
						\epsilon_{CP}=-\sum_j\frac{3}{2}\frac{M_{\Sigma_1}}{M_{\Sigma_j}}\frac{
							\Gamma_{\Sigma_j}}{M_{\Sigma_j}}I_j\frac{V_j-2S_j}{3},\label{ecp}
					\end{eqnarray}
					where the $S_j, V_j$ and $I_j$ arises due to loop and vertex correction . They can be expressed as follows,
					\begin{eqnarray}
						S_j=\frac{M^2_{\Sigma_j}\Delta m_{ij}^2}{(\Delta m_{ij}^2)^2+M_{\Sigma_1}^2\Gamma_{\Sigma_j}^2},\quad V_j=2\frac{M_{\Sigma_j}^2}{M_{\Sigma_1}^2}\Big[\Big(1+\frac{M_{\Sigma_{j}^2}}{M_{\Sigma_1}^2}\Big)\text{Log}\Big(1+\frac{M_{\Sigma_{j}^2}}{M_{\Sigma_1}^2}\Big)-1\Big],
					\end{eqnarray}
					and 
					\begin{eqnarray}
						I_j=\frac{\text{Im}[(Y_{\Sigma}Y_{\Sigma}^\dagger)^2_{1j}]}{|Y_\Sigma Y_\Sigma|_{11}|Y_\Sigma Y_\Sigma^\dagger|_{jj}}, \quad \Delta m_{ij}^2=M^2_{\Sigma_j}-M_{\Sigma_i}^2.
					\end{eqnarray}
					The gauge interaction decoupling temperature can be estimated from the ratio of the decay width $\Gamma_{GB}$ to the Hubble rate $H$ as \cite{Hambye:2012fh},
					\be
					\frac{\Gamma_{GB}}{H}=\frac{\gamma_{GB}}{n_\Sigma^{\rm eq}H}\leq1,
					\ee
					where $\gamma_{GB}$ is the gauge interaction density normalized by the equilibrium triplet number density $n_\Sigma^{\rm eq}$, thus, in typical cases where inverse decays ($\ell_{R}H\leftrightarrow \Sigma$) are alive, the $B-L$ asymmetry will be generated when gauge interactions are decoupled at lower temperatures after the inverse decays are turned off. Conversely, suppose the inverse decays are decoupled, the CP-violating out-of-equilibrium decay of the triplet fermions will produce a sufficient $B-L$ asymmetry. We have considered a viable region of Yukawa couplings which does satisfy all the necessary theoretical and experimental bounds and matches our results from LFV. 
					In the lower triplet mass region, the gauge processes dominate the Yukawa processes, and the triplet abundance is diluted by gauge boson-mediated annihilation processes. Therefore, the flavour effects in this low mass region are effectively small and can be neglected safely \cite{AristizabalSierra:2010mv}.
					\subsection{Numerical approach}					
					TeV triplets are thermalized by gauge boson-mediated annihilation up to $z>>1$. The generation of the $B-L$ asymmetry, in that case, proceeds basically above this $z$ once the relic fraction that survives annihilation starts decaying. Sphaleron interactions transform this asymmetry into a $B$ asymmetry up to temperatures $T_{\rm dec}$ at which their
					reactions are suddenly decoupled by the spontaneous breaking of the $SU(2)$ symmetry \cite{Burnier:2005hp}.
					This constraint combined with $Y_{\rm BAU} \sim 10^{-11}$ implies the bound $M_\Sigma\ge1.6$ TeV \cite{Strumia:2008cf}.  This bound on mass is however not followed by the standard singlet fermion resonant leptogenesis framework in which the singlet can explain baryogenesis with mass below the TeV scale \cite{Pilaftsis:2003gt,Dev:2014laa,BhupalDev:2014oar,Deppisch:2010fr,DeSimone:2007edo,Das:2020vca}. One get the reason for this as, in the standard singlet decay case, the efficiency is determined by $\tilde{m}$, on the contrary in the fermionic triplet scenario, there is a dependence on $M_{\Sigma_{i}}$ that strongly suppresses the efficiency when $M_\Sigma \sim \mathcal{O}$(TeV).
					Therefore, in this work, we hold a safer ground consistent with the triplet fermion mass around 1.7 TeV, with a reasonable choice of Yukawa coupling that also satisfy light neutrino mass bounds. To execute sufficiently enhanced lepton asymmetry, we have chosen the masses in such a way that it satisfies $M_{\Sigma_2}-M_{\Sigma_1}\simeq\Gamma_{\Sigma_1}$ and we get an asymmetry produced which is $\epsilon_{CP}=0.23$.
					
					The Boltzmann equation best describe the dynamics of a system and, it would be convenient for us to write down the relevant BEs and solve them numerically to study the evolution pattern.
					In the fermion triplet case, different triplet components can be involved in the same gauge scattering processes, hence it would be convenient to use a single Boltzmann equation summing over all the triplet components. The Boltzmann equation for the evolution of the triplet fermion and $B-L$ are\footnote{All the $\Sigma$ that appear in this section is the lightest $Z_2$ even triplet fermion. Since the triplet masses are nearly degenerate, considering any $Z_2$ even fermion triplet here would not disturb the analysis.} \cite{AristizabalSierra:2010mv}:
					\bea
					\frac{dY_{\Sigma}}{dz}=&&-\frac{1}{s H z}\Bigg[\Big(\frac{Y_{\Sigma}}{Y_{\Sigma}^{\rm eq}}-1\Big)\gamma_{D}-2\Big(\frac{Y^2_{\Sigma}}{(Y_{\Sigma}^{\rm eq})^2}-1\Big)\gamma_{A}\Bigg],\\
					\frac{dY_{B-L}}{dz}=&&-\frac{1}{s H z}\Bigg[\Big(\frac{Y_{\Sigma}}{Y_{\Sigma}^{\rm eq}}-1\Big)\epsilon_{\Sigma}-\frac{Y_{B-L}}{2Y_l^{\rm eq}}(1+4\gamma_{\Sigma}^{scatt})\Bigg]\gamma_{D}.
					\eea
					where $z=M_{\Sigma}/T,H=1.66\sqrt{g_*}T^2/M_{Planck}$ is the Hubble constant, $\gamma_{A,D}$ stands for annihilation and decay reaction densities respectively. $\gamma_\Sigma^{scatt}$ is the scattering density arises from lepton number violating scattering processes, such as $\ell\ell\leftrightarrow H^*H^*$ and $\ell H\leftrightarrow$ $\bar{\ell} H^*$. The equilibrium number densities can be expressed as $Y_{\Sigma}^{\rm eq}=\frac{135g_s}{16\pi^4g_*}x^2K_2(z)$ and $Y_{\ell_{R}}^{\rm eq}=\frac{135 \zeta(3)g_s}{8\pi^4g_*}$. In these expressions, $K_2$ is the modified Bessel function of the second kind and $g_s=2$ is the respective internal degree of freedom for the fermions. The reaction densities for the gauge decay and annihilation processes are expressed as,
					\begin{eqnarray}
						\gamma_D=&&sY_{\Sigma}^{\rm eq}\Gamma_{\Sigma}\frac{K_1(z)}{K_2(z)},\\
						\gamma_A=&&\frac{M_{\Sigma}T^3}{32\pi^3}exp(-2z)\Big[\frac{111g^4}{8\pi}+\frac{3}{2z}\big(\frac{111g^4}{8\pi}+\frac{51g^4}{16\pi})+\mathcal{O}(1/z)^2\Big].
					\end{eqnarray}  
					Here, $g$ is the gauge coupling. One can notice that in the absence of the $\gamma_A$ term, the efficiency is almost the same as the type-I seesaw leptogenesis \cite{Pilaftsis:2003gt,Dev:2014laa,BhupalDev:2014oar,Deppisch:2010fr}, the only difference is by a factor of 3. The same thing also holds for inverse decay since the lepton decay has three times more probability of encountering a Higgs particle to produce a heavy triplet. The produced lepton asymmetry is then converted into baryon asymmetry of the Universe via the sphaleron transition, which can be parametrized with the number of fermion generation ($n_f$) and the number of Higgs doublets ($n_H$) as:
					\begin{eqnarray}
						Y_{BAU}=-\Big(\frac{8n_f+4n_H}{22n_f+13n_H}\Big)Y_{B-L}.
					\end{eqnarray}
					\begin{figure}[h]
						\includegraphics[scale=0.65]{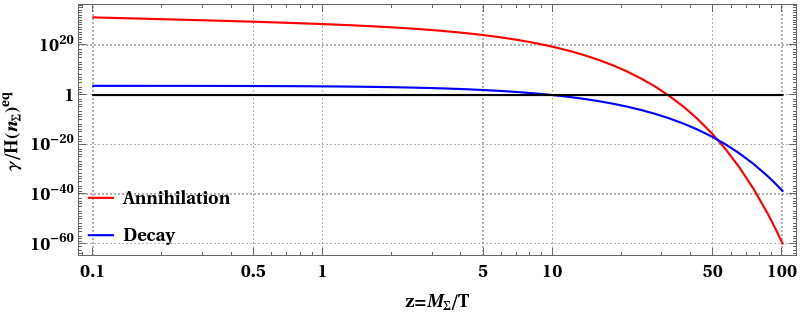}
						\caption{Ratio of annihilation(red) and decay(blue) rates with respect to Hubble rate. The fermion triplets $\Sigma_{1,2}$ decouple at a very early time.}\label{ratio1}
					\end{figure}
					\begin{figure}[h]
						\includegraphics[scale=0.5]{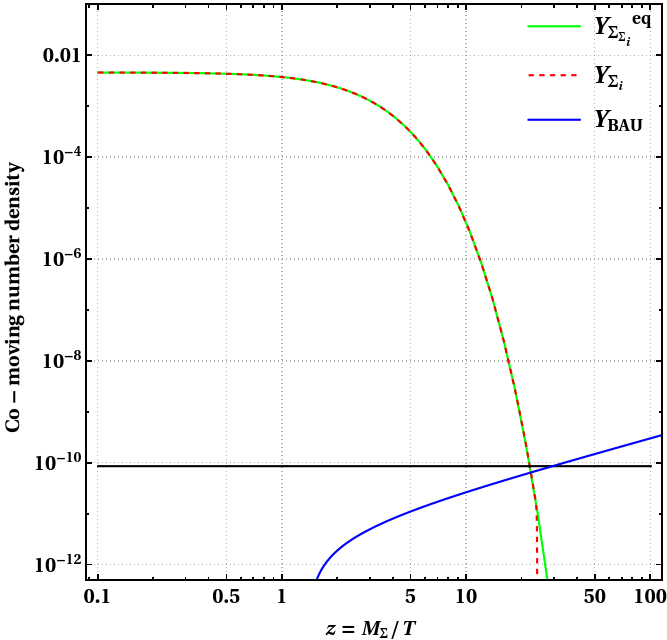}
						\caption{Variation of comoving number density $wrt$ $z=M_{\Sigma}/T$ for the fermion triplet itself (red dashed) and the observed baryon asymmetry of the Universe(blue). The black horizontal line corresponds to best fit value $Y_{\Delta}=8.75\times10^{-11}$.}
						\label{fig:lep1}
					\end{figure}
					
					We check the decoupling scenario in Fig. \ref{ratio1}, and one can see from here that the annihilation rates go out of equilibrium after decay rates. Due to the choice of triplet masses ($M_{\Sigma_1}\sim M_{\Sigma_2}\sim1.7$ TeV), the annihilation processes mediated via gauge processes are delaying the decay of $Z_2$ even triplets. The resonant scenario, where the choice of triplet mass does satisfy the light neutrino mass bounds, allows the production of asymmetry only after $z>15-20$ because, below this scale, the sphaleron decoupling forbids any sizable production of asymmetry. We can see from Fig. \ref{fig:lep1} that the observed baryon asymmetry reaches its current value for $z\sim 30$, just after the decay processes go out of equilibrium and the comoving density of the triplet fermion starts deviating from its equilibrium path. One can notice the baryon asymmetry production is reasonably delayed in this case and saturates at higher $z$ values.
					\section{Conclusion and Discussion}
					\label{sec:concl}
					
					In this work, we have extended the Standard Model by three hyperchargeless $Y=0$ real vector-like triplet fermions, among which two are $Z_2$ odd. We also added a $Z_2$-odd complex scalar triplet with hypercharge $Y=2$ to complete the model framework.
					This framework successfully explains the neutrino masses, dark matter, baryon asymmetry, inflation, and the reheating temperature of the Universe. 
					This unifying framework with fermion and scalar triplets is not in the literature, enabling us to present this work in detail.
					The two $Z_2$ even fermions can explain all the neutrino low energy variables at the TeV scale only using Casas-Ibarra parametrization~\cite{Goswami:2018jar}.
					The contributions to the neutrino mass from the scalar triplet at tree-level is zero as the $\bar{L} \Delta L$ term is absent due to the addition $Z_2$ symmetry transformation of $\Delta$. We also checked that $Z_2$ odd triplet can provide exact relic density through the freeze-out mechanism depending on the parameters in both the scalar and fermion ($\Sigma_3$).
					However, almost all the dark matter mass region for the neutral component of the scalar triplet is ruled out from the present direct detection constraints.
					The neutral component of the $Z_2$ odd fermion is slightly lighter than the charge component; hence we found that it gives a sizeable effective annihilation cross-section due to the co-annihilation channels. We get a smaller relic density (however allowed from the present data) for the low mass region $100-2000$ GeV. We get the exact relic density $\Omega h^2=0.1198 \pm 0.0012$ for dark matter masses of range $2.285<M_{DM}<2.445$ TeV, which is almost independent of the other parameters in this model. Furthermore, if the scalar triplet masses become closer to the $Z_2$ odd neutral fermion masses, another co-annihilation channel changes the dark matter parameter space, which is explained in detail.
					
					We then consider the CP-even neutral component of the scalar triplet as an inflaton and explain inflationary parameters like tensor-to-scalar ratio, spectral index, running spectral index, and scalar power spectrum.
					We have found the parameter space consistent with the latest Planck 2018 data. We also did the reheating analysis through the decay of the inflaton into the gauge and Higgs bosons.
					The decay and annihilation of these gauge and Higgs bosons (depending on the number densities) can produce relativistic lighter SM particles, which will reheat the Universe after the inflation. We have shown the parameter space, which gives the reheating temperature at $\mathcal{O}(10^{14})$ GeV.
					
					We also find the excess baryon asymmetry from the $Z_2$ even fermion triplets. The first two generations of $Z_2$ even fermions can successfully explain the neutrino parameters and the baryon asymmetry through resonant leptogenesis. In an alternative choice, the decay of the $Z_2$ odd triplet fermion field to the heavy Higgs fields and SM fermions can give additional CP-violation, which then contributes to the baryon asymmetry through leptogenesis. In contrast, the neutral component of the $Z_2$ odd fermion fails to present a viable dark matter candidate. Since the $Z_2$ odd triplet fermion leptogenesis does not fit with the current scenario (where $\Sigma_3$ is a DM candidate), we also present this alternative calculation in appendix~\ref{apndx}. 
					\section{Acknowledgements}
					NK would like to acknowledge support from the DAE, Government of India, for the Regional Centre for Accelerator-based Particle Physics (RECAPP), HRI. PD would like to acknowledge IITG for the financial support under the project grant number: IITG/R$\&$ D/IPDF/2021-22/20210911916. The authors also thank Anish Ghosal for useful discussions.
					
					\appendix
					\section{}
					\subsection{ An alternative approach: What if $\Sigma_3$ fails as a dark matter candidate?}\label{apndx}
					
					In this case, since only one triplet fermion is there, we can directly consider the loop and vertex correction terms from eq. \eqref{ecp} to be unity, $S_j=V_j=1$ ($\ell_{Ri}\, \Delta \leftrightarrow \Sigma_{3i}$).
					
					We here discuss this scenario for $M_{\Sigma}>M_{\Delta}$ with $\Delta=H,A,H^{\pm}$ and $H^{\pm\pm}$, although it is excluded from the dark matter point of view. At current scenario, we assume that dark matter may have some other origin.	We have estimated the CP asymmetry from the decay $\Sigma_{3}\rightarrow \ell_{Ri}\, \Delta$. This case is much simpler than the other triplet leptogenesis scenarios as the CP asymmetry takes the Universal form and is independent of the heavy state triplet mass. Decay process is solely dictated by the Yukawa coupling associated with the triplet. 
					The decay width of neutral component of the triplet has consequently the same decay width as one can see in the type-I leptogenesis scenario\cite{Hambye:2012fh, Davidson:2008bu, Buchmuller:2004nz,Das:2019ntw,Mahanta:2019gfe},
					\begin{equation}
						\Gamma_{\Sigma^0_3}=\Gamma(\Sigma^0_3\rightarrow \ell_{R}\Delta^+)+\Gamma(\Sigma^0_3\rightarrow \bar{\ell_{R}}\bar{\Delta}^+)=\frac{1}{8\pi}M_{\Sigma_3}|Y_{3i}|^2.\label{decay3}
					\end{equation}
					For the charged states, the right-handed and the left-handed conjugate states of $\Sigma_3^\pm$ form a Dirac spinor $\Psi^\pm$ and from the $SU(2)_L$ invariance, all the decay widths are the same as~\eqref{decay3}. Interestingly, the CP asymmetry generated from all the triplet states turned out to be three times less than the singlet fermion decay case\cite{Hambye:2012fh}. Hence, it is necessary to multiply the final lepton asymmetry by a factor of three in this case. The lepton asymmetry produced is given by \cite{Hambye:2012fh,Fischler:2008xm},
					\be 
					Y_{\Sigma_3}=	\frac{n_L}{s}=\epsilon_{\Sigma_3}\eta\frac{n_{\Sigma_3}}{s}|_{T>>M_{\Sigma_3}},
					\ee
					where $\epsilon_{\Sigma_3}$ is the CP asymmetry produced, $\eta$ is the efficiency, $n_{\Sigma_3}$ is the total number of triplets, including the particle and antiparticle, and $s$ is the entropy density expressed as $s=g_* (2\pi^2/45)T^3$ with $g_*$ being the relativistic degree of freedom whose value is 106.75 before electroweak phase transition. Although the triplet decay process is analogous to the singlet RH neutrino decay processes, some modifications cannot be ignored in the triplet scenario, such as gauge interactions\footnote{The fermion triplet scenario are relatively more straightforward in comparison to the scalar triplet decay. Apart from the gauge interactions, various decay and scattering processes influence the lepton asymmetry process.}. 					
					\begin{figure}
						\includegraphics[scale=0.6]{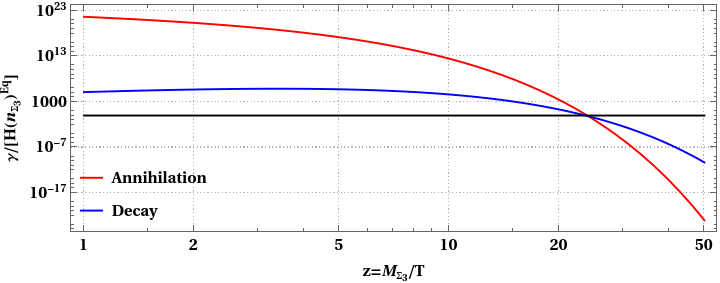}
						\caption{Decay and annihilation thermalization rates for the triplet fermion $\Sigma_3$. }\label{rate2}
					\end{figure}
					
					\begin{figure}
						\includegraphics[scale=0.45]{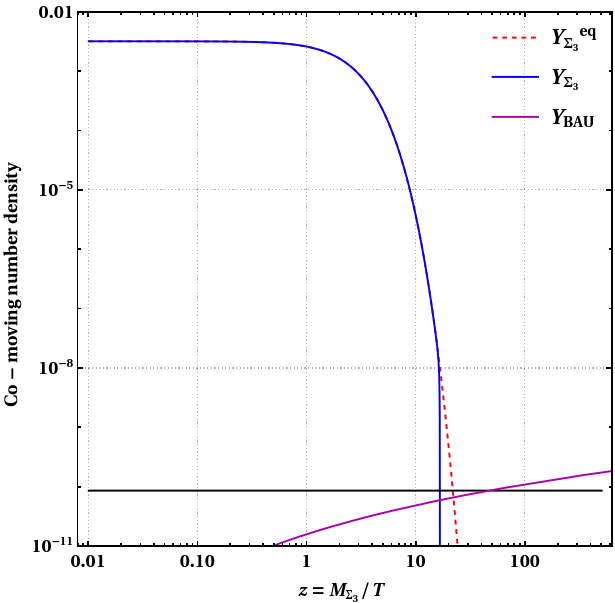}
						\caption{Variation of co-moving number density with $z=M_{\Sigma_3}/T$. The horizontal black line indicates the current value of comoving number density of baryons. }\label{leps3}
					\end{figure}
					We can see from Fig. \ref{rate2}, that the decay and annihilation processes thermalize approximately at the same time, hence this case can resemble the typical singlet fermion decay case. From Fig. \ref{leps3}, we can see that the triplet ($\Sigma_3$) number density deviates from the equilibrium number density (red dotted line) around $z\sim20$ and the observed baryon asymmetry value reaches its current value around $z\sim50$. The asymmetry production reaches saturation at a very late time, however, it does not get overproduced.
					
					This alternative choice of triplet fermion decay leptogenesis could play an interesting role in the dark matter parameter space. As we can see from the DM analysis from section \ref{sec:DM}, the Yukawa coupling associated with the $Z_2$ odd sector plays a vital role in producing DM relic freeze-out and, the same Yukawa couplings also mimic this leptogenesis scenario. In a situation where the neutral component from the scalar triplet ($\Delta$) behaves as a dark matter candidate, and the same Yukawa associated with the $Z_2$ odd sector will also influence this triplet fermion leptogenesis process. However, we do not include this choice in the present work, as this may confuse the readers; therefore, we keep this window open for future work. On a final note, we keep this alternative approach as a consequential part of this study to test {\it what if $\Sigma_3$ fails as a dark matter candidate?}

\section{Renormalization Group equations}
\label{sec:app}
In this study, we use the SM RGEs up to three loops which
can be found in Refs.~\cite{Chetyrkin:2012rz,Zoller:2013mra,Zoller:2012cv,Chetyrkin:2013wya}. The new field contributions are taken up to two loops which have been generated using SARAH~\cite{Staub:2015kfa}.
In this model, the RGEs of the couplings are defined as 
\bea
\beta_{\chi_{i}}=\frac{\partial \chi_{i}}{\partial \ln \mu} =   \frac{1}{16 \pi^2}~\beta_{\chi_{i}}^{(1)}  +  \frac{1}{(16 \pi^2)^2}~\beta_{\chi_{i}}^{(2)}\, .\nn 
\eea

The RGEs of the scalar quartic couplings $\lambda_{1,2,3,4,5}$ and Yukawa couplings upto one-loop are given by
\vspace{-0.5cm}
\subsection{Gauge Couplings}
\vspace{-1.2cm}
{\allowdisplaybreaks  \begin{align} 
\beta_{{g}_1}^{(1)} & =  \frac{47 g_1^3}{6} \\
\beta_{g_2}^{(1)} & =  \frac{3 g_2^3}{2}\\
\beta_{g_3}^{(1)} & =  
-7 g_{3}^{3} 
\end{align}} 

\vspace{-1.3cm}
\subsection{Quartic scalar couplings}
\vspace{-1.3cm}
{\allowdisplaybreaks  \begin{align} 
\beta_{ \lambda_1}^{(1)} & =  -3 g_1^2 \lambda _1-9 g_2^2 \lambda _1+\frac{3 g_1^4}{8}+\frac{3}{4} g_2^2 g_1^2+\frac{9 g_2^4}{8}+24 \lambda _1^2+\frac{5 \lambda _4^2}{4}\nonumber\\
   &~~~~~+3 \lambda _5^2+3 \lambda _4 \lambda _5+12 \lambda _1 y_t^2-6 y_t^4+6 \lambda _1 Y_{\Sigma }^2-\frac{5 Y_{\Sigma }^4}{2}\\
\beta_{ \lambda_2}^{(1)} & =  -\frac{15}{2} g_1^2 \lambda _4-\frac{33}{2} g_2^2 \lambda _4-12 g_1^2 g_2^2+4 \lambda _4^2+4 \lambda _1 \lambda _4+8 \lambda _2 \lambda _4\nonumber\\
   &~~~~~+4 \lambda _3 \lambda _4+8 \lambda _4 \lambda _5+6 \lambda _4 y_t^2+3 \lambda _4 Y_{\Sigma }^2+2 \lambda _4 Y_{{3i}}^2\\
\beta_{ \lambda_3}^{(1)} & =  -12 g_1^2 \left(g_2^2+\lambda _3\right)-24 g_2^2 \lambda _3+6 g_1^4+15 g_2^4+6 \lambda _2^2+28 \lambda _3^2\nonumber\\
   &~~~~~+2 \lambda _5^2+24 \lambda _2 \lambda _3+2 \lambda _4 \lambda _5+4 \lambda _3 Y_{{3i}}^2-2 Y_{{3i}}^4\\
\beta_{ \lambda_4}^{(1)} & =  -\frac{15}{2} g_1^2 \lambda _4-\frac{33}{2} g_2^2 \lambda _4-12 g_1^2 g_2^2+4 \lambda _4^2+4 \lambda _1 \lambda _4\nonumber\\
   &~~~~~+8 \lambda _2 \lambda _4+4 \lambda _3 \lambda _4+8 \lambda _4 \lambda _5+6 \lambda _4 y_t^2+3 \lambda _4 Y_{\Sigma }^2+2 \lambda _4 Y_{3i}^2\\
\beta_{ \lambda_5}^{(1)} & =  -\frac{15}{2} g_1^2 \lambda _5-\frac{33}{2} g_2^2 \lambda _5+3 g_1^4+6 g_2^2 g_1^2+6 g_2^4+\lambda _4^2\nonumber\\
   &~~~~~+4 \lambda _5^2+4 \lambda _1 \lambda _4+2 \lambda _2 \lambda _4+6 \lambda _3 \lambda _4+12 \lambda _1 \lambda _5\\
   &~~~~~+12 \lambda _2 \lambda _5+16 \lambda _3 \lambda _5+6 \lambda _5 y_t^2+3 \lambda _5 Y_{\Sigma }^2+2 \lambda _5 Y_{{3i}}^2\nonumber
\end{align}} 
\vspace{-1.3cm}
\subsection{Yukawa Couplings}
\vspace{-1.3cm}

{\allowdisplaybreaks  \begin{align}  
 \beta_{ y_t}^{(1)} & =  y_t \left(-\frac{1}{12} 17 g_1^2-8 g_3^2-\frac{9 g_2^2}{4}+3 y_t^2+\frac{3 Y_{\Sigma }^2}{2}\right)+\frac{3 y_t^3}{2}\\
 \beta_{ Y_{\Sigma} }^{(1)} & = Y_{\Sigma } \left(-\frac{1}{4} 3 g_1^2-\frac{33 g_2^2}{4}+3 y_t^2+\frac{3 Y_{\Sigma }^2}{2}\right)+\frac{5 Y_{\Sigma }^3}{4}\\
 \beta_{ Y_{3i} }^{(1)} & =Y_{3i} \left(-3 g_1^2-6 g_2^2+3 Y_{{3i}}^2\right)
\end{align}} 


					\bibliographystyle{apsrev4-1}
					\bibliographystyle{utphys}
					\bibliography{REFc}
					
	\end{document}